\documentclass[review]{elsarticle}

\usepackage{lineno,hyperref}
\usepackage{amsmath,amssymb,amsfonts}
\usepackage{algorithmic}
\usepackage{graphicx}
\usepackage{textcomp}
\usepackage{xcolor}

\journal{Optics \& Laser Technology}

\begin{document}

\begin{frontmatter}

\title{Influence of oceanic turbulence on propagation of autofocusing Airy beam with power exponential phase vortex }


\author[mymainaddress]{ Junzhe Wang}
\ead{q16010108@njupt.edu.cn}

\author[mymainaddress]{ Xinguang Wang}
\ead{xg-cgb@njupt.edu.cn}


\author[mymainaddress]{ Shengmei Zhao}
\ead{zhaosm@njupt.edu.cn}


\address[mymainaddress]{Institute of Signal Processing and Transmission, University of Posts and Telecommunications(NUPT), Nanjing, China}

\begin{abstract}
According to Rytov approximation theory, we derive the analytical expression of the detection probability of the autofocusing Airy beam (AAB) with power-exponent-phase carrying orbital angular momentum (OAM) mode, AAB-PEPV. We analyze the influence of oceanic turbulence on the propagation characteristics of the AAB-PEPV. The results show that the AAB-PEPV beam has a higher detection probability at the receiver when the anisotropic ocean turbulence has a larger unit mass fluid dynamic energy dissipation rate, a larger internal ratio factor, and a higher anisotropy factor. At the same time, the detection probability decreases with the temperature change dissipation rate, the temperature and salinity contribution to the refractive index spectrum. In addition, the larger power exponential phase and the longer wavelength the AAB-PEPV beam has, the better anti-interference the AAB-PEPV beam has. 
\end{abstract}

\begin{keyword}
Airy vortex beam \sep Orbital angular momentum\sep Propagation property \sep Oceanic turbulence

\end{keyword}

\end{frontmatter}


\section{Introduction}
As the significant increase in demand for data from underwater optical wireless communications (UOC),  underwater imaging and underwater sensor networks, the propagation characteristics of beams carrying orbital angular momentum (OAM) have attracted increasing attentions. Although many works have reported the feasibility of using OAM in UOC system to greatly increase the capacity by spatially multiplexing simultaneous data streams in a single transmitter/receiver pair \cite{baghdady2016blue,baghdady2016multi,ren2016orbital}, however, different OAM-carried beams have different anti-interference properties in the underwater channel, such as, partially coherent anomalous hollow vortex beam \cite{Liu2019}, Lorentz-Gauss beam \cite{Liu2018}, and partially coherent modified Bessel correlated vortex beam \cite{Chen2019}, etc. Here, OAM is a special phase structure of light beam, and has found a promising degree of freedom and been already studied in the quantum and free-space optical communications \cite{ zhao2017demonstration, Zhao2012}. 
Therefore, it is necessary to find the best candidate of the OAM-carried beam in underwater channel.

It is already demonstrated that the non-diffraction and self-healing OAM-carried beams, also named vortex beams,  have a better resistance to the distortion caused by turbulence, such as OAM-carried Bessel beam \cite{Zhao2019}, OAM-carried Airy beam \cite{Wang2019}. Since autofocusing Airy beam (AAB), described by
a radially symmetric or circular Airy function, could undergo abrupt autofocusing in free space \cite{Efremidis2010}, that is, the energy could suddenly increase right before the focal
point, many applications have been reported on practical trapping due to this property \cite{Jiang2013,Panagiotopoulos2013}. Furthermore, the AABs with noncanonical
optical vortex was reported to present some new optical characteristics \cite{Jiang2012}, they was demonstrated to be used as a steering parameter of optical vortex \cite{Kim2003}.
Since the propagation dynamics of
an optical vortex was influenced by the phase function itself, the one with a power exponential phase vortex, named  AAB-PEPV, was experimentally confirmed and discussed in \cite{li2014spiral}. 
Correspondingly, Lao\emph{ et. al.} applied a power exponential phase vortex to the Gaussian beam to study the intensity characteristics of this beam propagating in free space in \cite{lao2016propagation}. And Yan \emph{et. al.} analyzed the probability density of OAM mode of AAB-PEPV through weak anisotropic atmosphere
turbulence \cite{Yan2017}. HowIt is already demonstrated that the non-diffraction and self-healing OAM-carried beams have a better resistance to the distortion caused by turbulence, such as OAM-carried Bessel beam \cite{Zhao2019}, OAM-carried Airy beam \cite{Wang2019}. Since autofocusing Airy beam (AAB), described by
a radially symmetric or circular Airy function, could undergo abrupt autofocusing in free space \cite{Efremidis2010}, that is, the energy could suddenly increase right before the focal
point, many applications have been reported on practical trapping due to this property \cite{Jiang2013,Panagiotopoulos2013,Lu2019}. Furthermore, the AABs with noncanonical
optical vortex was reported to present some new optical characteristics \cite{Jiang2012}, they was demonstrated to be used as a steering parameter of optical vortex \cite{Kim2003}.
Since the propagation dynamics of
an optical vortex was influenced by the phase function itself, the one with a power exponential phase vortex, named  AAB-PEPV, was experimentally confirmed and discussed in \cite{li2014spiral}. 
Correspondingly, Lao\emph{ et. al.} applied a power exponential phase vortex to the Gaussian beam to study the intensity characteristics of this beam propagating in free space in \cite{lao2016propagation}. And Yan \emph{et. al.} analyzed the probability density of OAM mode of AAB-PEPV through weak anisotropic atmosphere
turbulence \cite{Yan2017}. However, to the best of our knowledge, the AAB-PEPV beam has not been investigated in anisotropic ocean turbulence.

In this paper, we study the propagation characteristics of AAB-PEPV beam through an anisotropic ocean turbulence. The analytical formula of the radial probability density of the OAM mode of the AAB-PEPV propagating through weak anisotropic oceanic turbulence is derived by Rytov theory. And the influence of different oceanic turbulence parameters and different beam parameters on  the AAB-PEPV beam are presented.

\section{The detection probability of OAM mode of AAB-PEPA beam through ocean turbulence}
In this section, we derive the detection probability of OAM mode of the AAB-PEPV beam through the weak anisotropic oceanic turbulence.
	
In the cylindrical coordinates $(r,\varphi ,z)$, the optical field of the AAB-PEPV beam at light source plane is  expressed as \cite{li2014spiral}
	\begin{equation}\label{eq:1}
	E^{(0)}(r,\varphi ,z=0)=Ai(\frac{r_{0}-r}{\omega })e^{(a\frac{r_{0}-r}{\omega })}e^{[2\pi i\upsilon (\frac{\varphi }{2\pi })^{n}]},
	\end{equation}
where $0 \leq\varphi \leq 2\pi$, $Ai(\cdot )$ represents the Airy function, $r_{0}$ and $\omega$ represent the radius and width of the AAB beam, $a$ is an exponential truncation factor (ranging from 0 to 1), $i$ denotes an imaginary unit, $\upsilon$ denotes the topological charge of the vortex, and $n$ denotes the power of the spiral phase, which can take both integers and fractions.
When the beam is transmitted through the oceanic channel, according to the Rytov theory\cite{andrews2005laser}, the optical field at distance $z$ is
	\begin{equation}\label{eq:2}
	E(\rho ,\theta ,z)=E_{p}(\rho ,\theta ,z)e^{[\psi (\rho ,\theta ,z)]},
	\end{equation}
where $\psi (\rho ,\theta ,z)$ is the complex phase interference factor to describe the interference caused by oceanic turbulence, $E_{p}(\rho ,\theta ,z)$ is the light field of the AAB beam transmitted to $z$  in free space without turbulence. Under paraxial approximation, $E(\rho ,\theta ,z)$ can be expressed as,
\begin{equation}\label{eq:3}
	\begin{aligned}
	E(\rho ,\theta ,z)&=-\frac{ik}{2\pi z}exp(ikz)\int \int drd\varphi \cdot E^{(0)}(r,\varphi ,z=0)\\
	&\times e^{\left \{ \frac{ik[\rho ^{2}+r^{2}-2\rho rcos(\theta -\varphi )]}{2z}+\psi (\rho ,\theta ,z)) \right \}},
	\end{aligned}
\end{equation}
where $k=\frac{2\pi}{\lambda}$ is wave vector, $\lambda$ is wavelength.
	
There is some difficulty to calculate the diffraction integral of Eq.(\ref{eq:3}) with Eq.(\ref{eq:1}) accurately. Here, we use a delta-ring of amplitude $A_{0}$ to represent this integral more accurately proposed in \cite{zhang2011trapping}, $E^{(0)}(r,\varphi ,z=0)\approx A_{0}\delta (r_{0}-r)$.
In this case, at $\rho $ = 0, the amplitude $A_{0}$ should be derived as 
	\begin{equation}\label{eq:4}
	A_{0}\approx \omega e^{(\frac{a^{3}}{3})}(1-\frac{\omega a^{2}}{r_{0}}).
	\end{equation}
where $\omega$ represents the width of the AAB-PEPV beam, and $a$ is the truncation factor.
Together with the properties of the first class Bessel function $J_l(\cdot)$, 
	\begin{equation}\label{eq:5}
	\begin{aligned}
	exp[\frac{z}{2}(t-\frac{1}{t})]&=\sum_{l=-\infty }^{\infty }J_{l}(z)t^{z},
	\\J_{-l}(z)&=(-1)^{l}J_{l}(z),
	\end{aligned}
	\end{equation}
 the analytical expression of the optical field of AAB-PEPV beam at the receiver plane can be obtained as
	\begin{equation}\label{eq:6}
	\begin{aligned}
	E(\rho ,\theta ,z)&=-\frac{ik}{2\pi z}e^{(ikz)}\omega (r_{0}\omega a^{2})e^{(\frac{ik\rho ^{2}+ikr_{0}^{2}}{2z}+\frac{a^{3}}{3})}\\
	&\times \left [  \sum_{l=-\infty }^{\infty }(-i)^{l}e^{(-il\theta )}M_{l}J_{l}(\frac{kr_{0}\rho }{z})\right ]e^{\left [ \psi (\rho ,\theta ,z) \right ]},
	\end{aligned}
	\end{equation}
	where
	\begin{equation}\label{eq:7}
	\\M_{l}=\int_{0}^{2\pi }e^{\left [ i\left ( 2\upsilon \pi \left ( \frac{\varphi }{2\pi } \right )^{n}+l\varphi  \right ) \right ]}d\varphi.
	\end{equation}
	
The result show that after propagation through the weak oceanic turbulence, the AAB-PEPV beam with a given topological charge $\upsilon$ is a weighted superposition of diffident OAM modes ($e^{(il\theta)}$) with $l$ varies from $-\infty$ to $\infty$. It is indicated that the fluctuation caused by oceanic turbulence has disturbed the OAM mode of ABB-PEPV beam. That is, the beam can not keep its original quantum state, and has the crosstalk to the adjacent OAM mode. Since the disturbance caused by oceanic turbulence is random, we analyze the average probability density distribution of the transmitted AAB-PEPV beam at the receiver plane.
	
It is known that the light field at the receiving plane can be decomposed to the superposition of OAM modes, since the OAM modes with different integer topological charges are orthogonal \cite{molina2001management}.
	\begin{eqnarray}\label{eq:8}
	&E(\rho ,\theta ,z)=\frac{1}{\sqrt{2\pi }}\sum_{m}^{}\beta _{m}(\rho ,z)exp(im\theta ),  \\ \nonumber
    &\beta_{m}(\rho ,z)=\frac{1}{\sqrt{2\pi }}\int_{0}^{2\pi }E\left (\rho ,\theta, z \right )\cdot exp(-im\theta )d\theta.
	\end{eqnarray}
%

Hence, the detection probability of the $m$ OAM mode inside the receiving end light field is given as
	\begin{equation}\label{eq:13}
\begin{aligned}
	\left \langle \left | \beta _{m}(\rho ,z) \right |^{2} \right \rangle=\frac{1}{2\pi }\int_{0}^{2\pi }\int_{0}^{2\pi }\left \langle E\left ( \rho , \theta ,z \right )E^{\ast }\left ( {\rho}' ,{\theta}' ,z \right )\right \rangle  exp\left [ im(\theta -{\theta}') \right ] d\theta d{\theta}',
\end{aligned}
	\end{equation}
where $\left \langle \cdot \right \rangle $ is the ensemble average, and $()^\ast$ is the complex conjugate. When the beam is transmitted in free space, the normalized correlation coefficient (i.e, the degree of coherence) at distance $z$ is consistent with that at the source ($z = 0$). Based on the statistical independence of the source and ocean turbulence, the cross-spectral density at transmission location $z$ should be
	\begin{equation}\label{eq:14}
\begin{aligned}
	W\left ( \rho ,{\rho}',\theta, {\theta}', z \right )\simeq \left \langle E\left ( \rho , \theta ,z \right )E^{\ast }\left ( {\rho}' ,{\theta}' ,z \right ) \right \rangle_{ot} \mu \left (\rho ,{\rho}',\theta, {\theta}'  \right ),
\end{aligned}
	\end{equation}
	where $\left \langle \cdot \right \rangle_{ot}$ is the ensemble average, $\mu \left (\rho ,{\rho}',\theta ,{\theta}' \right)$ is the coherence of the Airy-OAM \cite{morris2009propagation,zhang2011orbital}
	\begin{equation}\label{eq:15}
	\mu \left(\rho ,{\rho}',\theta ,{\theta}'\right )=e^{\left [-\frac{\rho ^{2}+{\rho }'^{2}-2\rho {\rho }'cos\left ( {\theta }' -\theta \right )}{2\rho _{s}^{2}}\right]},  \rho _{s}> 0,
	\end{equation}
	where $\rho_{s}$ is the coherence radius at the source plane $z=0$. Therefore, Eq.(\ref{eq:14}) can be rewritten as
	\begin{equation}\label{eq:16}
	\begin{aligned}
	W\left ( \rho ,{\rho}',\theta, {\theta}', z \right )&=E_{0}\left ( \rho , \theta ,z \right )E_{0}^{\ast }\left ( {\rho}' ,{\theta}' ,z \right )\\
&\times \left \langle exp\left [\psi(\rho ,\theta ,z)+\psi^{\ast}({\rho }',{\theta}',z) \right] \right \rangle_{ot}\\
	&\times exp\left [-\frac{\rho ^{2}+{\rho }'^{2}-2\rho {\rho }'cos\left ( {\theta }' -\theta \right )}{2\rho _{s}^{2}}\right],
	\end{aligned}
	\end{equation}
	where $E_{0}\left( \rho , \theta ,z \right )$ represents the complex amplitude of the AAB-PEPV beam in the z-plane.
	
	With the quadratic approximation, the second term on Eq.(\ref{eq:16}) can be expressed as
	\begin{equation}\label{eq:17}
	\begin{aligned}
	\left \langle e^{\left [\psi(\rho ,\theta ,z)+\psi^{\ast}({\rho }',{\theta}',z) \right]} \right \rangle_{ot}&=e^{\left[-\frac{1}{2}D(\rho ,{\rho }',z)\right ]}\\
	&=e^{\left [-\frac{\rho ^{2}+{\rho }'^{2}-2\rho {\rho }'cos\left ( {\theta }' -\theta \right )}{\rho _{o}^{2}}\right]},
	\end{aligned}
	\end{equation}
	where $D(\rho ,{\rho }',z)$ is the wave structure function and $\rho_{o}$ is the spatial coherence radius of the spherical wave propagating in the turbulence. Hence,  
Eq.(\ref{eq:16}) can be further rewritten as
	\begin{equation}\label{eq:18}
	\begin{aligned}
	W(\rho ,{\rho }',\theta ,{\theta }',z)&=E_{0}\left ( \rho , \theta ,z \right )E_{0}^{\ast }\left ( {\rho}' ,{\theta}' ,z \right )\\
&\times exp\left [-\frac{\rho ^{2}+{\rho }'^{2}-2\rho {\rho }'cos\left ( {\theta }' -\theta \right )}{\rho _{o}^{2}}\right]\\
	&\times exp\left [-\frac{\rho ^{2}+{\rho }'^{2}-2\rho {\rho }'cos\left ( {\theta }' -\theta \right )}{2\rho _{s}^{2}}\right]\\
	&=E_{0}\left ( \rho , \theta ,z \right )E_{0}^{\ast }\left ( {\rho}' ,{\theta}' ,z \right )\\
&\times exp\left [-\frac{\rho ^{2}+{\rho }'^{2}-2\rho {\rho }'cos\left ( {\theta }' -\theta \right )}{\tilde{\rho _{o}^{2}}}\right],
	\end{aligned}
	\end{equation}
	where $\frac{1}{\tilde{\rho_{o}^{2}}}=\frac{1}{\rho_{o}^{2}}+\frac{1}{2\rho_{s}^{2}}$, and $\frac{1}{\tilde{\rho_{o}^{2}}}$ are the effective spatial coherence radius, 
and $D(\rho ,{\rho }',z)$ is 
	\begin{equation}\label{eq:19}
\begin{aligned}
	D(r,{r}',z)& =8\pi ^{2}k^{2}z  \int_{0}^{1}\int_{0}^{\infty }\kappa \Phi (k ,\xi )\left [ 1-J_{0}\left ( \kappa \left | r-{r}' \right | \right ) \right ]d\kappa d\xi \\
&=\frac{2\left | r-{r}' \right |^{2}}{\tilde{\rho_{o}^{2}}},
\end{aligned}
	\end{equation}
	where $J_{0}(\cdot )$ is the first class zero-order Bessel function, and $\Phi(k ,\xi )$ is the anisotropic ocean current spectrum. 
Here, the anisotropy is assumed only exist in the propagation direction of the AAB-PEPV beam. The expression of $\Phi(k ,\xi )$ \cite{Nikishov2000} is
	\begin{equation}\label{eq:20}
\begin{aligned}
	\Phi (k,\xi ) =0.388\times 10^{-8}\chi _{t}\zeta ^{2}\varepsilon ^{-\frac{1}{3}}k^{-\frac{11}{3}}\left [ 1+2.35\left ( k\eta  \right ) ^{\frac{2}{3}}\right ] \phi (\kappa ,\varpi ),
\end{aligned}
	\end{equation}
	where $k=\sqrt{k_{z}^{2}+\zeta^{2}k_{\rho }^{2}}$, $k_{\rho }=\sqrt{k_{x}^{2}+k_{y}^{2}}$, 
$\chi _{t}$ is the rate of dissipation of the temperature variable (from $10^{-10}K^{2}\cdot s^{-1}$ to $10^{-4}K^{2}\cdot s^{-1}$), $\zeta $ is the anisotropy coefficient,  $\varepsilon $ is the rate of kinetic energy dissipation per unit mass of fluid (from $10^{-10}m^{2}\cdot s^{-3}$ to $10^{-1}m^{2}\cdot s^{-3}$), $\eta $ is the internal scale factor of the ocean current, $\varpi$ is the ratio of temperature and salinity contribution to refractive index spectrum (from -5 to 0), and $\phi (\kappa ,\varpi )$ is expressed as,
	\begin{equation}\label{eq:21}
\begin{aligned}
	\phi (\kappa ,\xi )=exp ( -A_{T}\sigma +\varpi ^{-2}exp(-A_{s}\sigma )-2\varpi ^{-1}exp(-A_{Ts}\sigma )),
\end{aligned}
	\end{equation}
where $A_{T}=1.863\times 10^{-2}$ , $A_{s}=1.9\times 10^{-4}$ , $A_{Ts}=9.41\times 10^{-3}$ , $\sigma =8.284\cdot (k\eta )^{\frac{4}{3}}+12.978\cdot (k\eta )^{2}$.

By simple calculation,  $\tilde{\rho_{o}}$ can be achieved as
	\begin{equation}\label{eq:22}
\begin{aligned}
	\tilde{\rho _{o}}=[ 8.705\times 10^{-8}k^{2}(\varepsilon \eta )^{-\frac{1}{3}}\zeta ^{-2}\chi _{t}z\times \left ( 1-2.605\varpi ^{-1}+7.007\varpi ^{-2} \right ) ]^{-1}.
\end{aligned}
	\end{equation}
	
Therefore, the detection probability of the $m$ OAM mode at the receiving end can be derived as, 
	\begin{equation}\label{eq:24}
	\begin{aligned}
	&\left \langle \left | \beta _{m}(\rho ,z) \right |^{2} \right \rangle\\
&=\frac{1}{2\pi }\int_{0}^{2\pi }\int_{0}^{2\pi }\left \langle E\left ( \rho , \theta ,z \right )E^{\ast }\left ( {\rho}' ,{\theta}' ,z \right )\right \rangle e^{\left [ im(\theta -{\theta}') \right ]} d\theta d{\theta}'\\
	&=\frac{1}{2\pi }\int_{0}^{2\pi }\int_{0}^{2\pi }W(\rho ,{\rho }',\theta ,{\theta }',z)e^{\left [ im(\theta -{\theta}') \right ]} d\theta d{\theta}'\\
	&=\frac{1}{2\pi }\int_{0}^{2\pi }\int_{0}^{2\pi }E_{0}\left ( \rho , \theta ,z \right )E_{0}^{\ast }\left ( {\rho}' ,{\theta}' ,z \right ) e^{\left [ im(\theta -{\theta}') \right ]}\\
	&\times exp\left [-\frac{\rho ^{2}+{\rho }'^{2}-2\rho {\rho }'cos\left ( {\theta }' -\theta \right )}{2\tilde{\rho _{o}^{2}}}\right]d\theta d{\theta }'\\
	&=2\pi\left [ \left ( \frac{k\omega }{2\pi z}\right )^{2}\cdot (r_{0}-\omega a^{2})^{2} \cdot exp\left ( \frac{2a^{3}}{3}-\frac{2\rho ^{2}}{\tilde{\rho }_{o}^{2}} \right ) \right ]\\
& \times \left [ \sum_{l=-\infty }^{\infty } M_{l}M_{l}^{\ast }\cdot J_{l}\left ( \frac{kr_{0}\rho }{z} \right )^{2}\cdot I_{l-m}\left ( \frac{2\rho^{2}}{\tilde{\rho}_{o}^{2}} \right )\right ].
	\end{aligned}
	\end{equation}

For the specified OAM mode when $m=\upsilon $, $\left \langle \left | \beta_{m}(\rho ,z) \right |^{2} \right \rangle $ expresses the detection probability of $\upsilon$ OAM mode at the receiving end. The larger value of  $\left \langle \left | \beta_{m}(\rho ,z) \right |^{2} \right \rangle $ shows the the beam has a good anti-interference through the turbulence. 
Assuming $\Delta m=\left |\upsilon -m\right | $, and $ \Delta m > 0 $, then $\left \langle \left | \beta_{\Delta m}(\rho ,z) \right |^{2} \right \rangle $ indicates the crosstalk from $\upsilon$ to $m$, named it crosstalk probability, which implies the probability of the transmitted $\upsilon$ OAM mode transferring to $m$ OAM mode due to the disturbance. The smaller value of $\left \langle \left | \beta_{\Delta m}(\rho ,z) \right |^{2} \right \rangle $ hints a better anti-interference property of the beam.
	
Assumed that the topological charge of the transmitted AAB-PEPV beam is $m_{0}$. 
Since there is an aperture for the receiver, 
the detection probability $P_{m_{0}}$ of the transmitted AAB-PEPV beam with $m_{0}$ after the oceanic turbulence channel should be defined as
\begin{equation}\label{eq:25}
\begin{aligned}
	P_{m_{0}}&=\frac{E_{m_{0}}}{\sum _{m}E_{m}}= \frac{\int_{0}^{R}\left \langle \left | \beta_{m_0}(\rho ,z) \right |^{2} \right \rangle rdr}{\sum_{m}{\int_{0}^{R}\left \langle \left | \beta_{m}(\rho ,z) \right |^{2} \right \rangle rdr}} \\
& =\int_{0}^{R}2\pi\left [ \left ( \frac{k\omega }{2\pi z}\right )^{2}\cdot (r_{0}-\omega a^{2})^{2} \cdot exp\left ( \frac{2a^{3}}{3}-\frac{2\rho ^{2}}{\tilde{\rho }_{o}^{2}} \right ) \right ] \\
&\times \left [ \sum_{l=-\infty }^{\infty } M_{l}M_{l}^{\ast }\cdot J_{l}\left ( \frac{kr_{0}\rho }{z} \right )^{2}\cdot I_{l-m_{0}}\left ( \frac{2\rho^{2}}{\tilde{\rho}_{o}^{2}} \right )\right ]rdr\\
&\div \sum _{m}\int_{0}^{R}2\pi\left [ \left ( \frac{k\omega }{2\pi z}\right )^{2}\cdot (r_{0}-\omega a^{2})^{2} \cdot e^{\left ( \frac{2a^{3}}{3}-\frac{2\rho ^{2}}{\tilde{\rho }_{o}^{2}} \right )} \right ] \\
&\times \left [ \sum_{l=-\infty }^{\infty } M_{l}M_{l}^{\ast }\cdot J_{l}\left ( \frac{kr_{0}\rho }{z} \right )^{2}\cdot I_{l-m}\left ( \frac{2\rho^{2}}{\tilde{\rho}_{o}^{2}} \right )\right ]rdr.
\end{aligned}
\end{equation}

\section{Numerical simulation and discussion}
n this section, we analyze the propagation characteristics of the AAB-PEPV beam in the anisotropic ocean turbulence. We mainly discuss the detection probability varying with the beam parameters and the oceanic turbulence parameters. In addition, the other parameters are listed as the following, the topological charge $\upsilon=1$, wavelength $\lambda =417nm$, propagation distance $z=100m$ and $z=300m$ , the exponential truncation factor $a=5\times 10^{-2}$, beam main lobe width $\omega =0.01m$, beam main lobe radius $r_{0}=10^{-3}m$, the kinetic energy dissipation per unit mass of fluid rate $\varepsilon =10^{-5}m^{2}\cdot s^{-3}$, the dissipation of the temperature variable rate $\chi _{t}=10^{-8}K^{2}\cdot s^{-1}$, internal scale factor $\eta =10^{-3}m$, the anisotropy coefficient $\zeta =2$, the ratio of temperature and salinity contribution to refractive index spectrum $\varpi =-3$ and the receiving radius $R=0.05m$.
\begin{figure}[!htp]
		\centering
		\includegraphics[width=0.8\columnwidth]{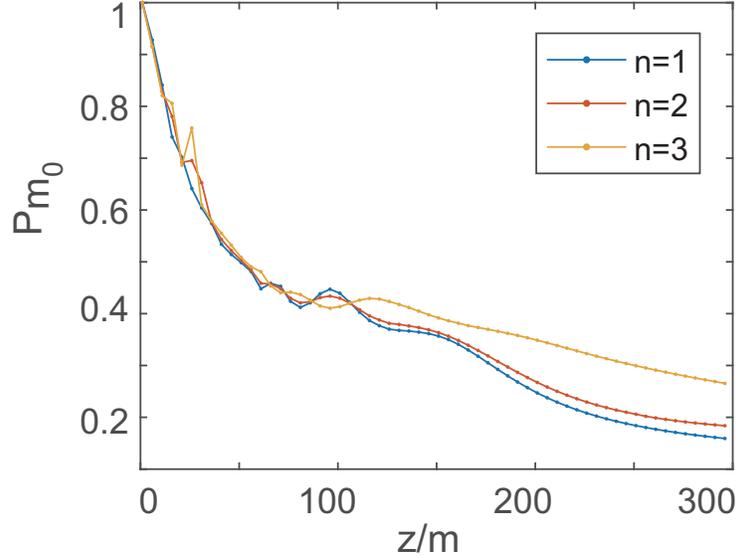}
		\caption{The detection probability $P_{m_{0}}$ against $z$ for different phase powers $n$ for the AAB-PEPV beam.}
		\label{FIG:1}
\end{figure}

Firstly, we analyze the effect of beam parameters on the detection probability. Here, the phase power of the beam $n$ is varied and the others are fixed. Figure \ref{FIG:1} shows the detection probability against $z$ for different phase powers $n$. The results show that the detection probability decreases with the propagation distance. Additionally, the detection probabilities of different phase powers have large fluctuations when the distance is less than $100m$, and the detection probability increases with the phase power when the propagation distance is greater than $100m$.  
Therefore, the AAB-PEPV beam  with a higher phase power is more conducive to the transmission in ocean turbulence.

	\begin{figure}[!htp]
		\centering
		\includegraphics[width=0.8\columnwidth]{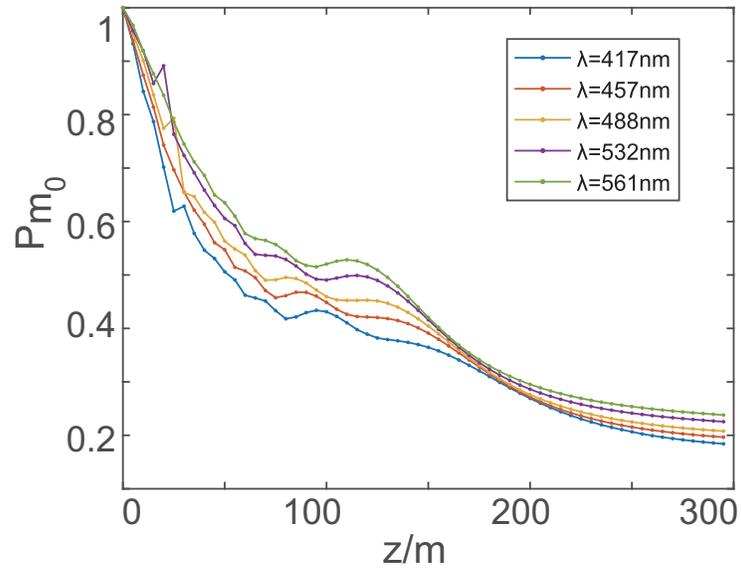}
		\caption{The detection probability $P_{m_{0}}$ against $z$ for different wavelength $\lambda$ of AAB-PEPV beam.}
		\label{FIG:2}
	\end{figure}

Secondly, we discuss the variation of the detection probability with the propagation distance $z$ when the wavelengths $\lambda$ are different. Figure \ref{FIG:2} shows the relationship between the detection probability $P_{m_{0}}$ and the transmission distance for different wavelengths. The results show that a slight fluctuation occurs for the shorter wavelength AAB-PEPV beam. This is because the detection probability in the radial direction increases first and then decreases as the wavelength increases at a shorter wavelength. At the same time, the shorter wavelength corresponds to a high wavenumber, resulting in a relatively strong flickering effect. 
Therefore, the longer wavelength of the AAB-PEPV beam is more helpful to the propagation through oceanic turbulence.

\begin{figure}[!htp]
		\centering
		\includegraphics[width=0.8\columnwidth]{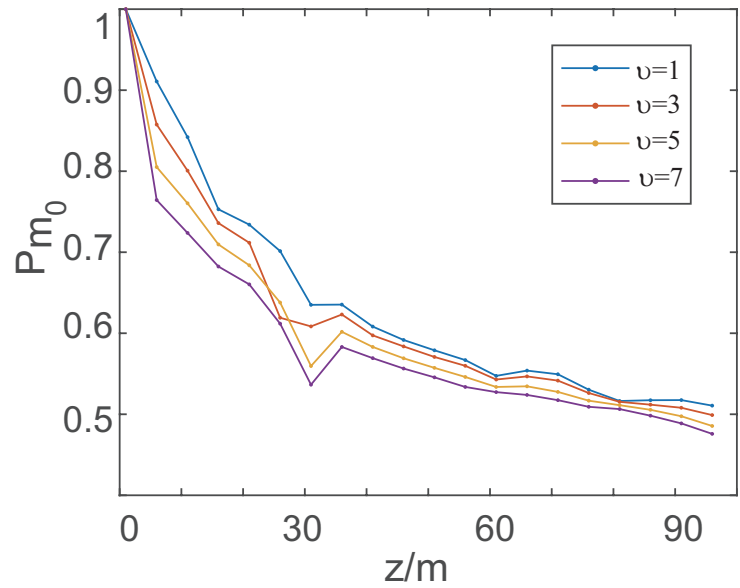}
		\caption{The detection probability $P_{m_{0}}$ against $z$ for different topological charge $\upsilon$ of the AAB-PEPV beam.}
		\label{FIG:3}
	\end{figure}

Thirdly, we demonstrate the detection probability of the AAB-PEPV beam against the transmission distance $z$ under different topological charges $\upsilon$ in Figure \ref{FIG:3}. 
The results show that the detection probability decreases as the topological charge increases for the same transmission distance. The AAB-PEPV beam with a smaller topological charge has a better anti-interference.

\begin{figure}[!htp]
 \centering
 \includegraphics[width=1.0\columnwidth]{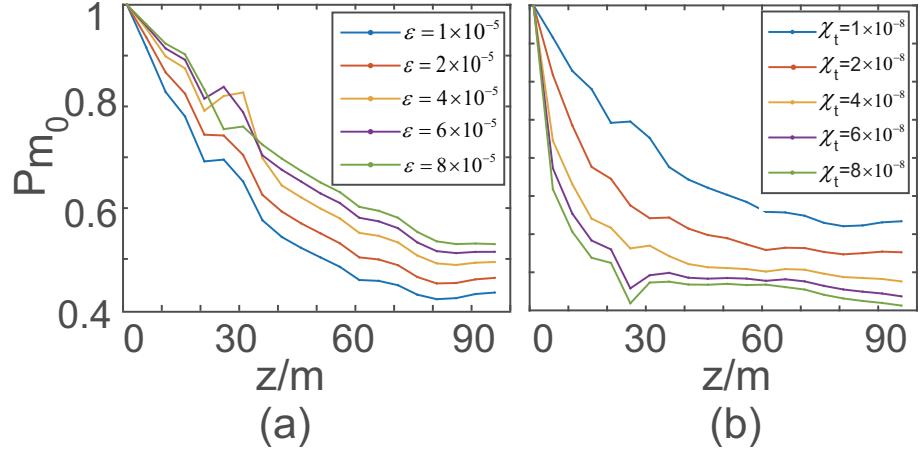}
 \caption{The detection probability $P_{m_{0}}$ against $z$ for different dynamic energy dissipation rates $\epsilon$ (a), and for different temperature dissipation coefficients $\chi _{t}$ (b) .}
 \label{FIG:4}
\end{figure}
	
Further, we analyze the influence of oceanic environment parameters on the detection probability. Figure \ref{FIG:4} shows the relationship between $P_{m{0}}$  and the transmission distance when the AAB-PEPV beam propagates through oceanic turbulence at different dynamic energy dissipation rates $\epsilon$ (see Fig.~\ref{FIG:4}(a)) and temperature dissipation coefficients $\chi _{t}$ (see Fig.~\ref{FIG:4}(b)). 
The results show that as the dynamic energy dissipation rate increases and the temperature dissipation coefficient decreases, the detection probability $P_{m{0}}$ increases. The larger the dynamic energy dissipation rate $\epsilon $ indicates the weaker oceanic turbulence, and has less interference on the propagating beam. 
	
\begin{figure}[!htp]
		\centering
		\includegraphics[width=.8\columnwidth]{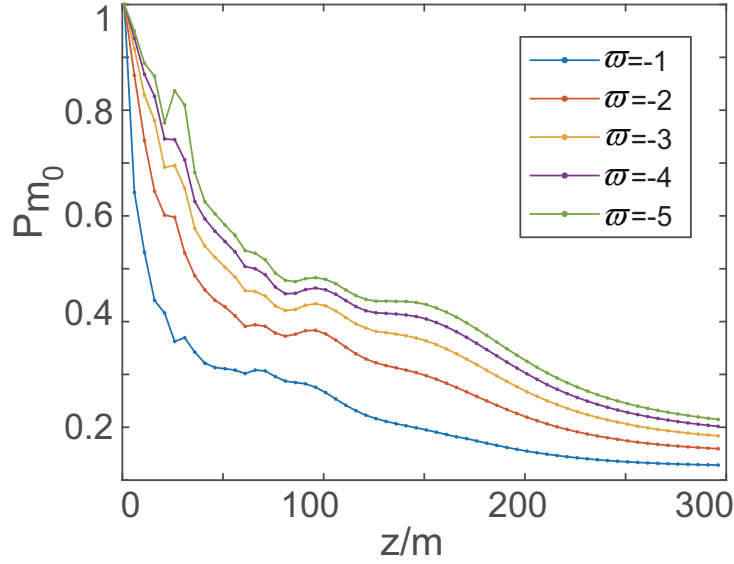}
		\caption{The detection probability $P_{m_{0}}$ against $z$ for different temperature and salinity contributions ratio $\varpi $ of Airy-OAM beam.}
		\label{FIG:5}
\end{figure}
At last, we present the influence of oceanic turbulence temperature and salinity contributions ratio to the refractive index spectrum $\varpi $ on the propagation characteristic of the AAB-PEPV beam in oceanic channel. Figure \ref{FIG:5} shows the detection probability $P_{m_{0}}$ against $z$ for different ratio of temperature and salinity contributions to the refractive index spectrum $\varpi $. It could be seen from the figure that $P_{m_{0}}$ decreases with the propagation distance $z$ for a given $\varpi $. For the same propagation distance, the smaller $\varpi $ is, the bigger the detection probability $P_{m_{0}}$ is.
The ratio of temperature and salinity contributions represents the effect of salinity and temperature on the transmission.

%


\section{Conclusion}
In this paper, we have demonstrated the propagation properties of the AAB-PEPV beam in an anisotropic weak oceanic turbulent channel. We have derived the analytic formula of detection probability of the AAB-PEPV beam after it propagating through the oceanic turbulence by using Rytov approximation theory. The influences of beam parameters, turbulence parameters, and propagation distance on the detection probability of the AAB-PEPV beam have been discussed. The results have shown that the disturbance caused by the oceanic turbulence on the propagation of the AAB-PEPV beam become stronger as the dissipation rate of temperature variance, the ratio of temperature and salinity contributions to the refractive index spectrum, and the propagation distance increase. Simultaneously, the detection probability of the AAB-PEPV beam after oceanic channel have decreased as the dissipation rate of kinetic energy per unit mass of fluid, inner scale factor, and anisotropic coefficient decrease. The AAB-PEPV beam with a large power exponential phase is more immune to oceanic turbulence. All the results are helpful for the UOC link using AAB-PEPV beam.

\section*{Acknowledgments}
The work is supported by Open Research Fund Program of the State Key Laboratory of Low-Dimensional Quantum Physics(KF201909); Partially supported by the National Natural Science Foundation of China (Grant No. 61871234);

\end{document}